\documentstyle[prd,aps,eqsecnum,preprint]{revtex}

\def\a{\alpha}
\def\b{\beta}

\def\e{\epsilon}
\def\g{\gamma}
\def\d{\delta}

\def\la{\lambda}
\def\L{\Lambda}
\def\th{\theta}
\def\Th{\Theta}
\def\om{\omega}
\def\del{\nabla}
\def\ha{\frac{1}{2}}
\def\bar#1{\overline{ #1 }}

\def\beq{\begin{equation}}
\def\eeq{\end{equation}}
\def\bea{\begin{eqnarray}}
\def\eea{\end{eqnarray}}
\def\bit{\begin{itemize}}
\def\eit{\end{itemize}}
\def\ben{\begin{enumerate}}
\def\een{\end{enumerate}}

\def\bdel{{\bf \nabla}}
\def\bA{{\bf A}}
\def\bB{{\bf B}}
\def\bD{{\bf D}}
\def\bal{{\bbox\alpha}}
\def\bx{{\bf x}}
\def\by{{\bf y}}
\def\bk{{\bf k}}
\def\bq{{\bf q}}
\def\bPi{{\bf \Pi}}
\def\dk{{d^3k\over(2\pi)^3}}
\def\dkn#1{\left(\prod_{n=1}^{#1} {d^3k_n\over(2\pi)^3}\right)}
\def\FP{{\cal J}}
\def\pri{\prime}
\def\Heff{H_{\rm eff}}
\def\Hcan{H_{\rm can}}
\def\Hct{H_{\rm CT}}
\def\ra{\rightarrow}

\begin{document}

\title{Renormalized Effective QCD Hamiltonian: Gluonic Sector}

\author{D. G. Robertson,$^1$ E. S. Swanson,$^{1,2}$ A. P. Szczepaniak,$^3$
C.-R. Ji,$^1$ and S. R. Cotanch$^1$}

\address{$^1$Department of Physics, North Carolina State University,
Raleigh, NC 27695-8202}

\address{$^2$Jefferson Laboratory, 12000 Jefferson Ave, Newport News, VA,
23606.}

\address{$^3$Department of Physics and Nuclear Theory Center, Indiana
University, Bloomington, IN 47405}


\maketitle

\begin{abstract}
Extending previous QCD Hamiltonian studies, we present a new
renormalization procedure which generates an effective Hamiltonian for
the gluon sector.  The formulation is in the Coulomb gauge where the
QCD Hamiltonian is renormalizable and the Gribov problem can be
resolved.  We utilize elements of the G{\l}azek and Wilson regularization
method but now introduce a continuous cut-off procedure which
eliminates non-local counterterms.  The effective Hamiltonian is
then derived to second order in the strong coupling
constant.  The resulting renormalized Hamiltonian provides a realistic
starting point for approximate many-body calculations of hadronic
properties for systems with explicit gluon degrees of freedom.
\\\\
\hspace{-1.5em}PAC number(s): 12.39.Mk, 12.39.Ki, 11.10.Ef, 11.10.Gh
\end{abstract}

\pacs{12.39.Mk, 12.39.Ki, 11.10.Ef, 11.10.Gh}

\section{Introduction}

A quantum field theory that is local will ensure that causality
is not violated, however, application of such a theory in
four-dimensional space-time will generate divergences.  About 50 years
ago in a series of papers \cite{dyson}, Dyson developed
renormalization to specifically address such infinites in QED.
He ingeniously converted the divergence problem into a useful
criterion for selecting appropriate theories, namely that they should
be renormalizable. Today the view towards renormalization is somewhat
more practical and perhaps limited in scope.  Guided by Wilson and
Weinberg, theories are now regarded more as partial descriptions
rather than complete constructs.  Accordingly, contemporary approaches
embody the concept of an effective
theory that only includes the
necessary degrees of freedom appropriate to a specific energy
range.


In this paper, we follow this philosophy and consider
the development of an effective QCD Hamiltonian for systems with
explicit gluon excitations in the few GeV energy range.  This work
extends our previous Hamiltonian glueball investigation \cite{ssjc96}
and recent renormalization treatment of the quark sector
\cite{ss97}. Our motivation is three-fold.  First, we seek a rigorous
formalism connecting QCD to effective and tractable model
Hamiltonians.  Second, we wish to develop a realistic theoretical
treatment to accurately describe hadron structure, especially systems
with gluonic degrees of freedom such as glueballs and hybrids.
Finally, we want to understand the basic confinement
and chiral symmetry breaking mechanisms.

We begin with the exact QCD Hamiltonian
formulated in the Coulomb gauge and focus upon the pure gluonic
sector.  Renormalizing the quark Hamiltonian, which has been previously
studied using a sharp cut-off regularization \cite{ss97}, and the
quark-glue sector will be addressed in a future publication.  There are
several reasons for utilizing the Coulomb gauge in which the divergence of
the color vector potential vanishes
($\bdel\cdot\bA=0$).
As detailed by Zwanziger \cite{zwanziger}, not only is
the Hamiltonian renormalizable in this gauge but the Gribov problem
\cite{gribov} can also be resolved.
The essence of the Gribov problem is that specification of $\bdel\cdot\bA=0$
does not uniquely fix the gauge in non-Abelian
gauge theories.  In general, there are many copies of each
gauge field configuration, all with the same divergence,
that are related by gauge transformations.  As the
true physical configuration space of a gauge theory is the set of
gauge potentials modulo local gauge transformations, one must select a
single representative from each set of gauge-equivalent
configurations.  The resulting sub-set, ${\cal A}_{\rm phys}$, of
independent field configurations is known as the fundamental modular region.

A convenient characterization of ${\cal A}_{\rm phys}$ is given by the
``minimal'' Coulomb gauge, obtained by minimizing a suitably-chosen
functional along gauge orbits.  We define the $L_2$ norm of a field
configuration along a gauge orbit by
\beq
F_\bA [G] = {\rm Tr}\int d^3x (\bA^G)^2\; ,
\eeq
where $G(\bx)$ is a gauge transformation and
\beq
\bA^G = G\bA G^\dagger - G\bdel G^\dagger\; .
\eeq
Then ${\cal A}_{\rm phys}$
is specified by choosing from each gauge orbit the
configuration which globally minimizes $F$.
It is straightforward to verify that the extrema of $F$ satisfy
$\bdel\cdot \bA=0$, so that this defines a Coulomb gauge.
Furthermore, at a minimum of $F$ the Faddeev-Popov operator satisfies
$M\equiv -\bdel\cdot \bD\geq0$ (i.e., its eigenvalues are
non-negative).  Here $\bD = \bdel-ig\bA$ is the covariant
derivative.  One can show that configurations for which $M=0$ occur
only on the boundary of ${\cal A}_{\rm phys}$.  These actually
represent gauge copies and must be identified to give the full
physical configuration space.  Needless to say, the resulting
functional space has an extremely complicated structure.  Zwanziger
has shown how to rigorously implement the restriction to ${\cal A}_{\rm
phys}$ in the infinite-volume limit for a gauge-fixed version of
the lattice QCD Hamiltonian formulation \cite{zwanziger}.

The confinement phenomenon in QCD has two complementary aspects: (1)
there is a long range attractive potential between colored sources;
(2)
the gluons which mediate this force are absent from
the spectrum of physical states.  This poses something of a mystery in
a covariant gauge, however, since for small $q^2$, (1) suggests that
the gluon
propagator should be more singular than $1/q^2$, while
(2) implies the propagator is suppressed.
Thus the mechanism for confinement is not
particularly transparent \cite{kondo}.

In the Coulomb gauge, in contrast, these two aspects can comfortably co-exist:
the long range force is represented by the instantaneous
Coulomb interaction, while the (transverse) gluon propagator
can simultaneously be suppressed at $q^2=0$.
Indeed, a detailed analysis suggests precisely this picture.  The key
issue is the proper identification of the
fundamental modular
region.  Zwanziger has
argued recently that restricting the configuration space to the
fundamental region results in the Coulomb term acquiring a singular
contribution at long distances \cite{zwanziger}.\footnote{This mechanism
for confinement was first suggested by Gribov \cite{gribov}.}
This singularity is also connected to a suppression of the propagator
for the would-be-physical transverse gluons at $q^2=0$.  This suggests a
rather appealing scenario for confinement in the Coulomb gauge.

For these reasons and because we are interested in
understanding confinement, we adopt the Coulomb gauge and only work
with transverse components of the gluon fields.  Next we divide the
Hamiltonian $H$ into a free part, $H_0$, which is the full Hamiltonian
evaluated with zero coupling constant, and interaction defined by
$H_I=H-H_0$.  We work in a Fock space spanned by eigenfunctions of
$H_0$ with eigenvalues $E_n$.  In general matrix elements of the
interaction diverge in this space and we regulate by
suppressing contributions between states in which the energy
difference $|E_n - E_m|$ is larger than scale set by a cut-off
parameter $\L$.  However, we do not use the sharp cut-off procedure
of G{\l}azek and Wilson \cite{gw93} since this generates unacceptable
non-local interactions
upon renormalization. These may be avoided by using a smooth regulator.
This interesting result is not
attributed to the specific Hamiltonian or choice of gauge since this
feature emerges in scalar field theories as well.  Although our theory
is now rendered finite it is inappropriately dependent upon the
parameter $\L$.  Furthermore, since this parameter is governed by free
energies,
the theory is no longer Lorentz invariant.  It is also not gauge
invariant but after renormalizing both symmetries will be restored
along with elimination of all the cut-off dependence.  In
addition to this regularization we also suppress divergent matrix elements
of one-body operators which arise from normal ordering the two-body
interaction. Again a continuous, exponential cut-off regulator is used and
cut-off sensitivity is removed by renormalization.

Renormalization is achieved by adding a counterterm Hamiltonian which
can be expressed in terms of unknown coefficients and a complete set
of local operators that respect the symmetries of the regularized
Hamiltonian.
An effective
and elegant means for determining these coefficients is by performing
similarity transformation\cite{gw93}.  Because the transformation is
unitary, the
physical content is preserved but the cut-off is now reduced to a
lower value $\L_1$.  The transformation also introduces new
interactions which incorporate the physics contained between the scales $\L$
and $\L_1$.  A key aspect of the renormalization scheme is the
requirement that the transformed Hamiltonian be form invariant, i.e.
it maintains its mathematical structure but with $\L_1$ now replacing $\L$
\cite{perry}.

Because $\L$ can be arbitrarily large, we
select an initial scale which is  amenable to perturbation
theory.  The regularized Hamiltonian is expanded in powers of the
coupling constant $g$ and for this study only terms to order
$g^2$ are retained.  While the similarity transformation
rigorously evolves the scale it cannot account for confinement at hadronic
energies. This constitutes our main theoretical omission.
For hadronic applications confinement may be described
by supplementing the renormalized Hamiltonian with a confining
interaction.  We will address this issue further in a future publication.

At this point we have obtained an effective Hamiltonian suitable for
application to the vacuum and excited gluonic states, i.e.
glueballs.  However, our experience \cite{ssjc96,orsay} indicates that an
improved description can be obtained by performing an
additional similarity transformation to a quasiparticle basis.  This
is a BCS (Bogoliubov) rotation which mixes the bare
parton gluon creation and annihilation operators.  As previously
demonstrated \cite{ssjc96,orsay} through variational calculations for
an unrenormalized Hamiltonian in both the quark and gluon sectors,
dynamical chiral symmetry breaking occurs and
reasonable values for condensates and constituent
masses are obtained.  Similar vacuum BCS variational calculations using
our renormalized
effective gluon Hamiltonian are in progress and will be reported in a
future publication along with many-body Tamm-Dancoff and Random Phase
Approximations for the excited hadron (glueball) spectrum.

This paper is divided into five sections and one appendix.  In the next
section our notation is established and the canonical QCD Hamiltonian
is specified.  Section III addresses our regularization scheme.
Section IV presents the main result of the paper. It includes a discussion of
the counterterm Hamiltonian and the form invariance renormalization scheme.
This is followed by a sub-section devoted to the
similarity transformation which yields the final renormalized
effective interaction.
A detailed representation of the canonical Hamiltonian in Fock space is
presented in the appendix.

\section{Canonical Hamiltonian}

Our starting point is the canonical QCD Hamiltonian in the Coulomb
gauge $\bdel\cdot\bA=0$ \cite{tdlee}.  We denote spatial
vectors by bold-faced quantities and use the matrix notation for the
gauge fields: $\bA\equiv \bA^a T^a$, with $T^a$ the generators of
SU($N_c$) in the fundamental representation, satisfying
$[T^a,T^b]=if^{abc}T^c$.  The gauge covariant derivative is $\bD =
\bdel-ig\bA$ operating on objects in the fundamental representation and
$\bD = \bdel-ig[\bA,\; \, ]$ for objects in the adjoint
representation.

The dynamical degrees of freedom are the transverse gauge fields
$\bA$, their conjugate momenta $\bPi$ (also transverse) and the quark
field $\psi$.  The canonical Hamiltonian takes the form
\beq
\Hcan = H_q + H_g + H_C + H_{qg}  ,
\label{breakup}
\eeq
with
\beq
H_q = \int d^3x \thinspace
\psi^\dagger(\bx)\left[-i\bal\cdot\bdel+\b m\right]\psi (\bx)
\eeq
\beq
H_g = {\rm Tr}\int d^3x\thinspace \left[\FP^{-1} \bPi \cdot
\FP \bPi + \bB\cdot\bB \right]
\eeq
\beq
H_C = \ha g^2 \int d^3x d^3y\thinspace
\FP^{-1}\rho^a(\bx) K^{ab}(\bx,\by) \FP \rho^b(\by)
\label{coulomb}
\eeq
\beq
H_{qg} = -g\int d^3x \psi^\dagger(\bx)\bal\cdot\bA(\bx)\psi(\bx)  .
\label{hqg}
\eeq
Here $\a^i\equiv\g^0\g^i$ and $\b\equiv\g^0$ are the Dirac
matrices, $\FP$ is the Faddeev-Popov determinant,
\beq
\FP = {\rm det}\left[\bdel\cdot\bD\right] ,
\eeq
normalized so that ${\rm det}\left[\bdel^2\right] = 1$, and
$\bB$ is the non-Abelian magnetic field with components
\beq
B_i^a = \e_{ijk}\del_jA^a_k + {g\over2}\e_{ijk}f^{abc}
A^b_jA^c_k  .
\eeq
In the Coulomb term, Eq. (\ref{coulomb}), the kernel $K$ is
represented in ``matrix'' notation as
\beq
K^{ab}(\bx,\by) = \langle\bx,a|(\bdel\cdot\bD)^{-1}(-\bdel^2)
(\bdel\cdot\bD)^{-1}|\by,b\rangle
\eeq
and the color charge density $\rho^a$ includes both
quark and gluonic contributions
\beq
\rho^a(\bx) = \psi^\dagger(\bx) T^a\psi(\bx)
+f^{abc} \bA^b(\bx) \cdot \bPi^c(\bx)  .
\eeq

The nonvanishing canonical commutator for the gauge field is
\beq
\left[A_i^a(\bx),\Pi^b_j(\by)\right]=
i\d^{ab}\left(\d_{ij}-{\del_i\del_j\over\bdel^2}\right)
\d^{(3)}(\bx-\by) .
\label{gluonccr}
\eeq
The Fourier decompositions are
\bea
A^a_i(\bx) &=& \int\dk {1\over\sqrt{2\om_\bk}}\left[
a_i^a(\bk) + {a^a_i}^\dagger(-\bk)\right]
e^{i\bk\cdot\bx}\\
\nonumber\\
\Pi^a_i(\bx) &=& -i \int\dk \sqrt{{\om_\bk\over2}}\left[
a_i^a(\bk) - {a^a_i}^\dagger(-\bk)\right]
e^{i\bk\cdot\bx}  ,
\eea
with $\om_\bk=|\bk|\equiv k$.
In momentum space Eq. (\ref{gluonccr}) reduces to
\beq
\left[a^a_i(\bk) , {a^b_j}^\dagger(\bk^\pri)\right]
=\d^{ab}(2\pi)^3\d^{(3)}(\bk - \bk^\pri)D_{ij}(\bk) ,
\eeq
where
\beq
D_{ij}(\bk) \equiv \d_{ij} - {k_i k_j\over \bk^2}.
\eeq
Note that transversality of $\bA$ and $\bPi$ requires
\beq
\bk\cdot{{\bf a}}^a(\bk) =
\bk\cdot{{\bf a}^a}^\dagger(\bk) = 0 .
\eeq
These operators can also be decomposed in terms of orthogonal polarization
vectors
\beq
a^a_i(\bk) = \sum_{\la =1,2}\epsilon_i(\bk,\la) a^a(\bk,\la) .
\eeq

For the Fermi field the operator expansion takes the form
\beq
\psi(\bx)=\sum_s\int\dk\left[ u(\bk,s)b(\bk,s)+
v(-\bk,s) d^\dagger(-\bk,s) \right] e^{i\bk\cdot\bx} ,
\eeq
where $s$ labels the helicity (color and flavor indices have been
suppressed) and a nonrelativistic normalization is used such that
$u^\dagger u=v^\dagger v=1$.
The nonvanishing canonical anticommutators are then
\beq
\left\{ b(\bk,s),b^\dagger(\bk^\pri,s^\pri)\right\} =
\left\{ d(\bk,s),d^\dagger(\bk^\pri,s^\pri)\right\} =
\d_{ss^\pri}(2\pi)^3\d^{(3)}(\bk - \bk^\pri) .
\eeq

It is straightforward to express the canonical
Hamiltonian in the Fock representation.  The result to ${\cal
O}(g^2)$ is given in the Appendix.

\section{Regularization Scheme}

Perturbative schemes for renormalizing Hamiltonians typically suffer from
technical problems related to the occurrence of vanishing energy
denominators.  An elegant way of avoiding this difficulty is the
cut-off method of G{\l}azek and Wilson \cite{gw93}.  This approach
uses the basis formed by eigenstates of the free Hamiltonian
$H_0$, with eigenvalues $E_n$.  The theory is then regulated by
suppressing matrix elements of $H$ between states for which the
difference in free energies $E_{nm}\equiv E_n-E_m$
is large.  Specifically, matrix elements
of the regulated Hamiltonian $H(\L)$ are defined by
\beq
\langle n|H(\L)|m \rangle \equiv E_n \d_{nm}
+f_{nm}(\L) \langle n| H_I |m\rangle ,
\eeq
where $H_I=H-H_0$ is the interaction.  Here
$f_{nm}(\L)$ is some convenient function which approaches unity for
$|E_{nm}|\ll \L$ and vanishes for $|E_{nm}|\gg \L$.  In this work we
employ a smooth cutoff of the form\footnote{We have
also investigated a sharp cut-off, with
$f_{nm}=\th(\L-|E_{nm}|)$.
However, this choice leads to pathologies in the
renormalized Hamiltonian, specifically nonlocal counterterms.  These terms
also arise in scalar field theory and thus have nothing to do with the choice
of the Coulomb gauge.  This issue is discussed further below.}
\beq
f_{nm}(\L) = e^{-E_{nm}^2/\L^2} .
\label{cutoff}
\eeq
Note that  diagonal matrix elements are unmodified.

A few remarks on this regularization scheme are in order.  First, it
does not reduce the size of the Hilbert space, that is, no states are
removed from the theory.  Reduction to a finite number of degrees of
freedom---which is necessary for the nonperturbative calculations we
wish eventually to perform---occurs later, after the variational
vacuum state is determined and the transition to constituent quarks
and gluons has been achieved.  At this point we can apply, for
example, a Tamm-Dancoff truncation and consider hadronic states built from
a small number of constituents.  This truncation can perhaps be
justified by the emergence of constituent-scale masses for the
effective degrees of freedom, which kinematically suppresses mixing with
more complex, multi-constituent states.  A similar truncation in the
bare parton basis (current quarks and mass zero gluons) would be essentially
meaningless.

Second, since the cut-off is defined in terms of {\em free} energy
differences it is not fully Lorentz invariant (though it is
rotationally invariant, of course).  This is unavoidable
in a Hamiltonian framework, particularly if positivity of the Hilbert space
norm is to be preserved (thus ruling out, e.g., the Pauli-Villars scheme).
Indeed we conjecture that requiring positivity forces one to employ
a noncovariant gauge.
The renormalized effective Hamiltonian will therefore contain
Lorentz-noninvariant operators, that is, operators which correspond to
Lorentz-noninvariant terms at the Lagrangian level.  The regulator also
violates gauge invariance, though this may be of less significance since we
work in a fixed gauge.  Nevertheless our renormalization procedure, to be
detailed below, will automatically provide the counterterms necessary to
restore both symmetries.

Finally, this scheme does not completely regulate the theory.
Normal ordering the two-body operators in the canonical Hamiltonian
leads to one-body operators (see Eqs. (\ref{1gluon}) and
(\ref{1coulomb}) in the appendix) with divergent matrix elements that
are not regulated by the above procedure.  For these matrix elements
we insert an additional cut-off
factor, $\exp(-2\om_{k^\pri}^2/\L^2)$, in the
integrands of Eqs. (\ref{1gluon}) and (\ref{1coulomb}).\footnote{Of
course, we could simply omit these operators from the
Hamiltonian since our goal is to remove the cut-off dependence by adding
counterterms.  However, the divergences from these terms cancel divergences
elsewhere in the theory so we retain them.}  While there is
substantial freedom in implementing this supplementary regularization,
observables computed from the Hamiltonian
will be independent of the cut-off prescription.  We remark
on this below.

At this point we have a fully regulated Hamiltonian formulation of QCD
in the Coulomb gauge.  The next task is to remove the dependence on
the cut-off parameter $\L$ by adding counterterms to the Hamiltonian.
This will be done perturbatively, which should be reasonable for QCD
if the cut-off is not too low.  The resulting
renormalized effective Hamiltonian, $\Heff(\L)$, can then be
analyzed nonperturbatively using many-body techniques.

\section{Cut-off Dependence and Renormalization}

\subsection{Counterterm Hamiltonian and Form Invariance}

Our goal is to construct the renormalized effective Hamiltonian,
$\Heff(\L)$, which contains a cut-off but gives
cut-off independent results through some fixed order in perturbation
theory.  It can be expressed as the sum of the canonical Hamiltonian and
a ``counterterm'' Hamiltonian
\beq
\Heff(\L) = \Hcan(\L) + \Hct(\L) .
\eeq
The canonical Hamiltonian contains
normal ordering contributions with integrals regulated as
previously described.  The counterterm Hamiltonian, which begins at order
$g^2$ in perturbation theory, has operators
with the same structure along with additional terms necessary to correct
for violations of Lorentz invariance introduced by the regulator.  Note
also that the matrix elements of $\Hct$ depend explicitly on $\L$ in
addition to containing the regulating exponentials.  In general,
\beq
\Hct(\L) = \sum_i c_i(\L) \L^{n_i} {\cal O}_i^\L ,
\label{genlhct}
\eeq
where the $\{{\cal O}_i^\L\}$ are a complete set of local operators
invariant under the symmetries preserved by the regulator.
The superscript $\L$ indicates that the regulating
functions are to be associated with the operators ${\cal O}_i$;
that is,
\beq
\langle n|{\cal O}_i^\L |m\rangle \equiv e^{-E_{nm}^2/\L^2}
\langle n|{\cal O}_i |m\rangle,
\eeq
where ${\cal O}_i$ is the ``bare'' operator.
The
explicit powers of $\L$ are inserted in Eq. (\ref{genlhct})
so that the coefficients
$c_i(\L)$ are dimensionless ($n_i$ is four minus the mass dimension of
the operator ${\cal O}_i$).  Operators for which $n_i<0$ are
irrelevant in the renormalization group sense and will generally be
ignored.
Note that the coefficients $c_i$ can depend on
$\L$ {\em only} through their dependence on the ``canonical''
couplings of the theory.
Thus
\beq
c_i(\L) = \tilde{c}_i g^2(\L) + \cdots ,
\eeq
where the ellipsis indicates higher order terms in the perturbative
expansion and for simplicity only a single independent coupling has
been assumed.  The pure numbers, $\tilde{c}_i$, are independent of
$\L$.
Thus the coefficients depend logarithmically on $\L$ and one may only
specify the
change in the coefficient.
This may be contrasted with, for example,  a gluon mass counterterm
which arises
solely because the regulator violates gauge invariance.  In this
case the gluon mass is completely calculable within perturbation
theory.




An exhaustive way of determining $\Hct$ would be to list all permissible
operators and then extract the $c_i$ by requiring observables
calculated perturbatively be both cut-off independent and Lorentz covariant.
A more appealing, physically equivalent, approach adopted here is
to perform a similarity transformation on the Hamiltonian which reduces
$\L$ to a new cut-off
$\L_1$ in the regulating exponentials.  Because the transformed
Hamiltonian is equivalent to the original one, calculated observables
(e.g., eigenvalues) are unchanged.  In addition to replacing
$\L\ra\L_1$ in $\Hcan$, the transformation also generates new, effective
interactions representing the physics between the scales $\L$ and
$\L_1$.  Schematically, we have
\beq
\Hcan(\L) \ra \Hcan(\L_1) + \d H(\L_1,\L)
\eeq
under the similarity transformation.  The counterterm Hamiltonian is
then determined by requiring that the full, transformed Hamiltonian have the
same {\em form} as the original one, but with $\L$ everywhere replaced
by $\L_1$.  This requirement of form invariance, or coherence, of the
Hamiltonian under changes of the cut-off ensures that we have
the most general Hamiltonian allowed \cite{perry,pw}.  In the language
of the renormalization group (RG), the self-similar Hamiltonian exists
on an infinitely long RG trajectory and thus constitutes a renormalized
Hamiltonian.

We now sketch this procedure to first order in
$\a_s\equiv g^2/4\pi$.  In this case  $\Hct$ is of order $\a_s$ and
since the
similarity transformation leaves $\Hct$ unchanged to this order (apart from
replacing $\L$ by $\L_1$ in just the regulating functions) we have
\beq
\Heff(\L)\ra \Hcan(\L_1) + \d H(\L_1,\L) + \Hct(\L) .
\eeq
In this expression $\Hct$ is understood to have $\L$ replaced
by $\L_1$ in the regulating functions but nowhere else.
We now demand that the transformed Hamiltonian be equal to
\beq
\Heff(\L_1) \equiv \Hcan(\L_1) + \Hct(\L_1).
\eeq
This uniquely determines the structure of $\Hct$.

A prescription for treating
the cut-off dependent terms from normal ordering must also be
specified since these are not regulated by the cut-off on energy differences
and are also unaffected by the similarity transformation.  Our
prescription is that upon performing the similarity transformation, we
replace $\L$ by $\L_1$ in these terms in the canonical Hamiltonian and
include the difference between this and the original terms---the ``slice''
coming from momenta between
$\L_1$ and
$\L$---in $\d H$.

We first determine contributions to $\d H$ coming from the
normal-ordering terms.  The one-body operators
in the effective Hamiltonian have the general form
\bea
\Heff^{\rm (2)}(\L) &=& \int\dk\left\{\left[\om_k\d_{ij}\d^{ab}
+\Pi^{ab}_{ij}(\bk)\right] {a^a_i}^\dagger(\bk) a^b_j(\bk)
\right.\nonumber\\
& &\left. \qquad\qquad +
M^{ab}_{ij}(\bk) e^{-(2\om_k)^2/\L^2} \left[a^a_i(\bk) a^b_j(-\bk)
+{\rm h.c.}\right] \right\} .
\eea
Note the presence of the regulating function in the
second term.  From rotational invariance,
\beq
\Pi^{ab}_{ij}(\bk) = A^{ab}(k)\d_{ij} + B^{ab}(k)k_ik_j,
\eeq
with similar structure for $M^{ab}_{ij}$.  Due to transversality, however,
only the term proportional to $\d_{ij}$ need be considered.
{}From Eqs. (\ref{1gluon}) and (\ref{1coulomb}) we can identify the
contributions to $\Pi^{ab}_{ij}$ and $M^{ab}_{ij}$ arising from normal
ordering the two-body operators.  From the Coulomb term we have
\beq
\Pi^{(C)ab}_{ij}(k) = {\a_s \over4}N_c \d^{ab}\int
{d^3k^\pri\over(2\pi)^3} \widetilde{V}(\bk+\bk^\pri)
\left({\om_k^2+\om_{k^\pri}^2\over\om_k\om_{k^\pri}}\right)
D_{ij}(\bk^\pri)e^{-2\om^2_{k^\pri}/\L^2}
\eeq
and
\beq
M^{(C)ab}_{ij}(k) = {\a_s\over8}N_c\d^{ab}
\int {d^3k^\pri\over(2\pi)^3} \widetilde{V}(\bk+\bk^\pri)
\left({\om_{k^\pri}^2 - \om_k^2\over\om_k\om_{k^\pri}}\right)
D_{ij}(\bk^\pri)e^{-2\om^2_{k^\pri}/\L^2} .
\eeq
Next the cut-off is reduced by an infinitesimal
amount, $\L^2\rightarrow\L_1^2$, with
\beq
\L_1^2 \equiv (1-\e)\L^2\; ;\qquad \e>0 .
\eeq
The difference
\beq
\d\Pi^{(C)ab}_{ij}(k;\L,\e)\equiv
\Pi^{(C)ab}_{ij}(k;\L) - \Pi^{(C)ab}_{ij}(k;\L_1)
\eeq
to be included in $\d H$ is then expanded to first order in $\e$, followed by
an expansion in powers of $k/\Lambda$.  Retaining only
those terms which correspond to relevant or marginal operators in the
renormalization group sense yields
\footnote{It may
be necessary to retain some irrelevant operators if the
cut-off is reduced much below the scale of physical
interest,
but for the present we ignore these.}
\beq
\d\Pi^{(C)ab}_{ij}(k;\L,\e) =
{\a_s\over4\pi}N_c\d_{ij}\d^{ab} \left({\e\over\om_k}\right)
\left[{1\over3}\L^2 + {16\over15}k^2+\cdots\right]  ,
\eeq
where the dots represent terms of order $k/\L$ and higher.
Similarly we obtain
\beq
\d M^{(C)ab}_{ij}(k;\L,\e) =
{\a_s\over4\pi}N_c\d_{ij}\d^{ab}
\left({\e\over\om_k}\right)
\left[{1\over6}\L^2 - {2\over15}k^2+\cdots\right]  .
\eeq
Performing the same steps for the divergent contributions
in $H_g$ [Eq. (\ref{1gluon})] produces
\beq
\d\Pi^{(g)ab}_{ij}(k;\L,\e) =
{\a_s\over4\pi} N_c \d_{ij}\d^{ab} \left({\e\over\om_k}\right)
\left({2\over3} \L^2+\cdots\right)
\eeq
and
\beq
\d M^{(g)ab}_{ij}(k;\L,\e) =
{\a_s\over4\pi} N_c \d_{ij}\d^{ab}
\left({\e\over\om_k}\right) \left({1\over3} \L^2+\cdots\right) .
\eeq

\subsection{Similarity Transformation}

We now develop the similarity transformation renormalization procedure.
This follows
the formulation of Wegner \cite{wegner} who derived flow equations for
diagonalizing a Hamiltonian.  Consider a continuous transformation
governed by a unitary operator $U$ which depends upon a (flow)
parameter $\lambda$.  Let $T(\lambda)$ be the generator of the corresponding
infinitesimal transformation.  Then for a continuous transformation from
$\lambda_i$ to $\lambda_f$
\begin{equation}
U(\lambda_f;\lambda_i) =
e^{\int^{\lambda_f}_{\lambda_i}{T(\lambda)d\lambda}},
\end{equation}
where $\lambda$-ordering of the exponential is understood.
Note that $T$ must be anti-hermitian since $U^{\dagger} =
U^{-1}$.  The similarity transformation for a Hamiltonian which also depends
on $\lambda$
is then given by
\begin{equation}
H(\lambda_f) = U(\lambda_f;\lambda_i) H(\lambda_i)
U^\dagger(\lambda_f;\lambda_i),
\label{simtrans}
\end{equation}
and for an infinitesimal evolution
\beq
\label{dhl}
{dH(\lambda)\over d\lambda} = [T(\lambda) , H(\lambda)] .
\eeq

Wegner has determined that the generator choice
\beq
\label{th}
T(\lambda) = [H_0,H(\lambda)] ,
\eeq
suppresses the off-diagonal matrix elements thereby
rendering $H$ ``more diagonal" as $\lambda$ increases.
We have independently obtained the same
generator choice
by requiring that the transformation
properly change the cut-off parameter in the regulating
functions for all matrix elements of our Hamiltonian.  We now
demonstrate this but first note, from Eqs. (\ref{dhl}, \ref{th}), the
dimensions of $\lambda$ are $[E^{-2}]$ so to work directly with our
energy cut-off parameter we substitute $\lambda \rightarrow
\Lambda^{-2}$.  Then combining the above equations yields

\beq
{dH(\L)\over d\L^{-2}} = \Bigl[[H_0,H(\L)],H(\L)\Bigr] .
\label{dhdlambda}
\eeq
By expanding the interaction Hamiltonian in powers of the coupling
at this scale $\L$,
\beq
H_I(\L) = \sum_{p=1}^\infty h_p ,
\eeq
we can examine Eq. (\ref{dhdlambda}) order by order.
At  zeroth order
\beq
{dH_0\over d\L^{-2}} = 0 ,
\eeq
so that the transformation preserves the cut-off independence of the
free Hamiltonian. At first order we find
\beq
{dh_1(\L)\over d\L^{-2}} = \Bigl[[H_0,h_1],H_0\Bigr] .
\eeq
Expressed in terms of matrix elements this becomes
\beq
{d\over d\L^{-2}}\langle n|h_1(\L)|m\rangle
= - E_{nm}^2 \langle n|h_1(\L)|m\rangle .
\label{ogsimtrans}
\eeq
Integrating Eq. (\ref{ogsimtrans}) from $\L$ to $\L_1$
then yields
\beq
\langle n|h_1(\L_1)|m\rangle =
e^{-E_{nm}^2\left(1/\L_1^2 - 1/\L^2\right)}
\langle n|h_1(\L)|m\rangle .
\eeq
Since
\beq
\langle n|h_1(\L)|m\rangle \propto e^{-E_{nm}^2/\L^2} ,
\eeq
we see the the transformation has indeed replaced $\L\ra\L_1$ in the
regulating exponentials for these terms.

The expression at second order reads
\bea
{d\over d\L^{-2}}\langle n|h_2(\L)|m\rangle
&=& - E_{nm}^2 \langle n|h_2(\L)|m\rangle \nonumber\\
& & +\sum_l(E_n+E_m-2E_l)
\langle n|h_1(\L)|l\rangle \langle l|h_1(\L)|m\rangle .
\label{og2}
\eea
Once again the first term on the right-hand side has the effect of
replacing $\L\ra\L_1$ in the overall regulating exponential.  It is
trivial to see that this same structure occurs for each order, so that
the transformation does indeed replace $\L\ra\L_1$ in all matrix
elements of the Hamiltonian.
It is convenient to isolate this explicit cut-off
dependence by defining ``reduced'' matrix elements, denoted
by $V_p$
\beq
\langle n|h_p(\L)|m\rangle \equiv
e^{-E_{nm}^2/\L^2} \langle n|V_p(\L)|m\rangle .
\eeq
Notice that $V_1$ does not depend on $\L$.
In terms of $V_2$, Eq. (\ref{og2}) reads
\beq
e^{-E_{nm}^2/\L^2}
{d\over d\L^{-2}}\langle n|V_2(\L)|m\rangle
=\sum_l(E_{nl}+E_{ml})
e^{-(E_{nl}^2+E_{ml}^2)/\L^2}
\langle n|V_1|l\rangle \langle l|V_1|m\rangle .
\eeq
This can be integrated to give
\beq
\langle n|\d V_2|m\rangle
= \sum_l\Th(E_{nl},E_{ml})
\langle n|V_1(\L)|l\rangle \langle l|V_1(\L)|m\rangle ,
\eeq
where
\beq
\langle n|\d V_2|m\rangle \equiv
\langle n|V_2(\L_1)|m\rangle - \langle n|V_2(\L)|m\rangle.
\eeq
This change in the interaction explicitly
represents the physics removed in reducing
the cut-off from $\L$ to $\L_1$, to this order in
perturbation theory.  The quantity
\beq
\Th(E_{nl},E_{ml})\equiv
\ha\left({1\over E_{nl}}+{1\over E_{ml}}\right)
\left(e^{-2E_{nl}E_{ml}/\L^2} - e^{-2E_{nl}E_{ml}/\L_1^2}\right)
\eeq
plays the role of an energy denominator in
conventional perturbation theory.  Note that cancellation between the
two exponentials suppresses the contribution when the energy
differences are small.  Thus the potentially singular factors
$1/E_{nl}$, etc., are not problematic.
It is straightforward to extend this procedure to higher orders.

We can now compute the various one-body operators induced by
the similarity transformation.  Consider first the
contribution to $\d\Pi$ arising from $q\bar q$ intermediate states.
It is given by
\beq
\d\Pi^{(quark,1)ab}_{ij}(k) = {g^2\d^{ab}\over4\om_k} \sum_{s,s^\pri}
\int{d^3q\over(2\pi)^3} \thinspace
\Th(\Delta,\Delta)
\left[u^\dagger(\bk+\bq,s)\a_i v(-\bq,s^\pri)\right]
\left[v^\dagger(-\bq,s^\pri)\a_j u(\bk+\bq,s)\right] ,
\eeq
where
\beq
\Delta\equiv \om_k-E_q-E_{k+q}.
\eeq
We again expand this for a small change in the cut-off and discard
contributions corresponding to irrelevant operators.
The result is
\beq
\d\Pi^{(quark,1)ab}_{ij}(k;\L,\e) = -{\a_s\over4\pi}
N_f \d_{ij}\d^{ab}
\left({\e\over\om_k}\right)\left({1\over12}\L^2+
{1\over6}\sqrt{{\pi\over2}}k\L+ \cdots\right) ,
\eeq
where the dots represent terms of order $1/\L$ and $N_f$ is the number
of light (i.e., lighter than the cut-off scale) quark flavors.
There is also a contribution arising from $q\bar q gg$ intermediate
states (analogous to a ``Z-graph'' contribution) which gives
\beq
\d\Pi^{(quark,2)ab}_{ij}(k;\L,\e) = -{\a_s\over4\pi}
N_f \d_{ij}\d^{ab}
\left({\e\over\om_k}\right)\left({1\over12} \L^2-
{1\over6}\sqrt{{\pi\over2}} k\L+ \cdots\right) .
\eeq
The complete contribution arising from multi-quark intermediate
states is thus
\beq
\d\Pi^{(quark)ab}_{ij}(k;\L,\e) = -{\a_s\over4\pi}
N_f \d_{ij}\d^{ab}
\left({\e\over\om_k}\right)\left({1\over6}\L^2+
\cdots\right).
\eeq

The contribution from multi-gluon intermediate states may be
evaluated similarly; the result is
\beq
\d\Pi^{(glue)ab}_{ij}(k;\L,\e) =
-{\a_s\over4\pi} N_c \d_{ij}\d^{ab} \left({\e\over\om_k}\right)
\left({1\over12} \L^2 + {16\over 15}k^2 +\cdots \right) .
\eeq
Combining these results with the contributions from the
normal-ordering terms gives the complete $\d\Pi$ at this order:
\beq
\d\Pi^{ab}_{ij}(k;\L,\e) =
{\a_s\over16\pi} \d_{ij}\d^{ab}
\left({\e\over\om_k}\right)
\left[\left(11-{2N_f\over3} \right)\L^2 + \cdots\right],
\label{fulldpi}
\eeq
where we have set $N_c=3$.  We recognize the lowest-order
coefficient of the QCD beta function $\b_0\equiv
11-2N_f/3$.\footnote{This is purely accidental since an alternative
regulator choice in the normal-ordering
terms gives a different result.}

Next we consider the various contributions to $\d M^{ab}_{ij}$.
The $q\bar qg$ intermediate states yields
\beq
\d M^{(quark)ab}_{ij}(k;\L,\e) =
-{\a_s\over4\pi} N_f \d_{ij}\d^{ab}
\left({\e\over\om_k}\right) \left({1\over12} \L^2+\cdots\right),
\eeq
while three-gluon intermediate states give
\beq
\d M^{(glue)ab}_{ij}(k;\L,\e) =
-{\a_s\over4\pi} N_c \d_{ij}\d^{ab}
\left({\e\over\om_k}\right) \left({1\over24} \L^2
+{8\over 15}k^2 +\cdots\right).
\eeq
Combining these results with the contributions from the
normal-ordering terms gives
\beq
\d M^{ab}_{ij}(k;\L,\e) =
{\a_s\over16\pi} \d_{ij}\d^{ab}\left({\e\over\om_k}\right)
\left[\ha\b_0 \L^2
- 8 k^2+\cdots\right] .
\label{fulldm}
\eeq
for $N_c=3$.

One-body operators in the counterterm Hamiltonian have the general form
\beq
\Hct(\L) = \int\dk\left(
\Pi_{\rm CT}(k;\L) {a^a_i}^\dagger(\bk) a^a_i(\bk)
+ M_{\rm CT}(k;\L) e^{-(2\om_k)^2/\L^2} \left[a^a_i(\bk) a^a_i(-\bk)
+{\rm h.c.}\right] \right) ,
\eeq
since the cut-off dependent terms are all
proportional to $\d_{ij}\d^{ab}$.  Let us focus first on the term in
$\Pi_{\rm CT}$
that depends quadratically on $\L$. On dimensional grounds one has
\beq
\Pi_{\rm CT} = {\a_s\over 4\pi}\left({1\over2\om_k}\right)
b\L^2 + \cdots ,
\label{basicstruct}
\eeq
where $b$ is a numerical constant independent of $\L$.
After the similarity transformation we have
\beq
\Hcan(\L)+\Hct(\L) \ra \Hcan(\L_1) + \d H(\L_1,\L) + \Hct(\L) ,
\label{trafo}
\eeq
where $\d H$ represents the contributions computed above.
Note that to this order, the only change to $\Hcan(\L)$ is to
replace $\L$ with $\L_1$.
Equating the second and third terms
on the RHS of Eq. (\ref{trafo}) to $\Hct(\L_1)$ yields, using Eq.
(\ref{basicstruct}),
\bea
\ha \e\b_0\L^2 + b\L^2 &=& b\L_1^2\nonumber\\
&=& b\L^2(1-\e) .
\eea
Thus $b=-\b_0/2$, and
\beq
\Hct(\L) = -{\a_s\b_0\L^2\over8\pi}\int\dk\left(1\over2\om_k\right)
{a^a_i}^\dagger(\bk) a^a_i(\bk) + \cdots .
\eeq
It may be verified that this counterterm
removes the quadratically divergent part of the gluon self energy to
second order, as it should.  Of course, this is simply (part of)
a gluon mass term with the mass given by
\beq
m_g^2 = -{\a_s\b_0\L^2\over8\pi} .
\eeq
Note that the quadratically divergent part of $\d M$ has exactly this
structure.  Therefore, when expressed in terms of the field
operators the complete quadratically divergent part of the
counterterm Hamiltonian is simply
\beq
\Hct(\L) = {m_g^2\over2}\int d^3x A^a_i(\bx) A^a_i(\bx) +\cdots .
\eeq

The remaining operators in $\Hct$ may be similarly constructed.
Dimensionally, the term in Eq. (\ref{fulldm}) containing the factor
$-8k^2$ represents a logarithmic divergence. Hence, it should
correspond to the only logarithmic divergence to which we are
sensitive at this order, namely the gluon wavefunction
renormalization.  To  see this we invert the
Fourier expansions and express this operator in terms of the fields to
obtain
\beq
\d H = -\e {\a_s\over4\pi} \left({4C_A\over3}\right)
{\rm Tr}\int d^3x \left(\bB^2 - \bPi^2\right) ,
\eeq
where we have restored the dependence on the number of colors
($C_A=N_c$ is the Casimir invariant of the adjoint representation of
SU($N_c$)).  This has precisely the correct structure for a
wavefunction renormalization since, if the gauge field is rescaled by
$A\ra Z_A^{1/2}A$ then the conjugate momentum is rescaled by the
inverse factor, $\Pi\ra Z_A^{-1/2}\Pi$. \footnote{A quick way to see
this is to note that in a functional Schr\"odinger representation
$\Pi=-i\d/\d A$.}

The counterterm Hamiltonian will thus contain a term
\beq
\Hct = c(\L){\rm Tr} \int d^3x \left(\bB^2 - \bPi^2\right)\vert_{g=0}
\eeq
at this order, with $c(\L)$ logarithmically dependent on $\L$.
The requirement of coherence of the Hamiltonian under a change of the
cut-off from $\L\ra\L_1$ then gives
\beq
-\e {\a_s\over4\pi}\left({4C_A\over3}\right) + c(\L) = c(\L_1).
\eeq
Expanding $c(\L_1)$ about $\L_1=\L$ yields
\beq
\L{d\over d\L} c(\L) = {\a_s\over4\pi}\left({8C_A\over3}\right)
\eeq
to ${\cal O}(\e)$ and hence
\beq
c(\L) = {\a_s\over4\pi}\left({8C_A\over3} \right)
\ln(\L/\L_0) ,
\eeq
where $\L_0$ is an arbitrary scale.

As discussed above, one can interpret this counterterm as corresponding
to a renormalization of the gauge field with form
\beq
A^\pri = \left[1+\ha c(\L)\right]A \equiv Z_A^{1/2}A .
\eeq
Thus, the free part of $\Hcan$
combines with this part of the counterterm Hamiltonian to simply produce
a free Hamiltonian when written in terms of $A^\pri$.
The gauge field
anomalous dimension is now easily computed to be
\bea
\g_A &=& -\ha\L{d\over d\L}\ln Z_A \nonumber\\
\nonumber\\
&=& -{g^2\over16\pi^2}\left({4C_A\over3}\right).
\eea
At lowest order, there is no contribution to $\g_A$ from
fermions in this scheme.

It is significant to note that with a sharp cut-off, the above results
are again obtained, but with two differences.  One is a slight
difference in the value of the gluon mass.  The other, which is
problematic, is a contribution
\beq
H_{NL} = \left({\a_s\over4\pi}\right)
\e\b_0\L \int {d^3k\over (2\pi)^3}
\left[a_i(\bk) a_i(-\bk) + a_i^\dagger(\bk)a_i^\dagger(-\bk)\right]
\label{nonlocalct}
\eeq
to the counterterm Hamiltonian.
This corresponds to a nonlocal interaction, as may easily
be seen by inverting the field expansions and writing
Eq. (\ref{nonlocalct})
in terms of the field operators
\beq
H_{NL} = -\left({\a_s\over4\pi}\right)\e\b_0\L
\int d^3x d^3y {1\over |\bx - \by|^2}\left[\bA(\bx)\cdot
\del^2 \bA(\by) + \bPi(\bx)\cdot \bPi(\by)\right]\;.
\label{nonlocal}
\eeq
The nonlocality arises because the interaction
kernel is now a polynomial in $|\bk|$ rather than $\bk$.



\section{Conclusions}

The main, new result of this study was the rigorous derivation of a
renormalized effective Hamiltonian to order $g^2$ in the strong
coupling constant. This was achieved through a similarity
transformation in conjunction with a continuous cut-off
regularization scheme.  The use of a sharp cut-off regularization led
to an unacceptable non-local Hamiltonian.  The effective Hamiltonian
was determined by introducing an
appropriate counterterm Hamiltonian and imposing  form invariance on
the transformed Hamiltonian.
Renormalization not only restored the proper
symmetries but also rendered all observables calculated with this
Hamiltonian insensitive to the cut-off parameter.

Future formal work will address both the issue of confinement as well
as renormalization in the combined quark-gluon sector.  In
particular both perturbative and non-perturbative approaches will be
examined.  Such a treatment is fundamental to understanding the role
of gluons in the structure of mesons and baryons.  Finally, utilizing
many-body techniques, large scale applications of the effective
Hamiltonian focusing upon glueball and hybrid meson
systems will be reported.

\acknowledgments

\noindent
It is a pleasure to thank
B. Allen,
G. P. Lepage
and D. Zwanziger
for several informative discussions.
This work was supported by DOE grants DE-FG02-96ER40944,
DE-FG02-97ER41048 and DE-FG02-96ER40947.

\appendix

\section*{Canonical Hamiltonian}

In this Appendix we present the canonical Hamiltonian expressed in the
Fock representation.
The free Hamiltonian is simply
\beq
H_0 = \int\dk\thinspace \om_k \thinspace
{a^a_i}^\dagger(\bk) a^a_i(\bk)
+\sum_s \int\dk\thinspace E_k \thinspace\left[
b^\dagger(\bk,s) b(\bk,s) + d^\dagger(\bk,s) d(\bk,s) \right]  ,
\eeq
where $E_k=\sqrt{\bk^2+m^2}$ and $\om_k = |\bk|$.
The interaction terms will be separated according to their term of
origin in Eq. (\ref{breakup}).  From the purely gluonic term, $H_g$,
we have three-point couplings to ${\cal O}(g)$
\bea
H_g^{(3)} &=& {ig\over2\sqrt{2}} f^{abc}\int\dkn3 {k_{1j}\over
\sqrt{\om_1\om_2\om_3}} (2\pi)^3\d^{(3)}(\sum_m \bk_m) \\
& &\times {\bf :}
\left[a^a_i(\bk_1) + {a^a_i}^\dagger(-\bk_1)\right]
\left[a^b_j(\bk_2) + {a^b_j}^\dagger(-\bk_2)\right]
\left[a^c_i(\bk_3) + {a^c_i}^\dagger(-\bk_3)\right]
{\bf :}  ,\nonumber
\eea
where $\om_1\equiv \om_{\bk_1}$, etc.

For ${\cal O}(g^2)$ we obtain (normal-ordered) four-point couplings
\bea
H_{g}^{(4)} & &  ={\a_s\pi\over4} f^{abc}f^{ade} \int\dkn4
{1\over\sqrt{\om_1\om_2\om_3\om_4}}
(2\pi)^3\d^{(3)}(\sum_m\bk_m)\label{2gluon}\\
& &\times {\bf :}
\left[ a^b_i(\bk_1) + {a^b_i}^\dagger(-\bk_1)\right]
\left[ a^c_j(\bk_2) + {a^c_j}^\dagger(-\bk_2)\right]
\left[ a^d_i(\bk_3) + {a^d_i}^\dagger(-\bk_3)\right]
\left[ a^e_j(\bk_4) + {a^e_j}^\dagger(-\bk_4)\right]
{\bf :} ,\nonumber
\eea
where $\a_s\equiv g^2/4\pi$.
In addition, there are one-body operators which arise in normal
ordering Eq. (\ref{2gluon})
\bea
H_g^{(2)} &=& \a_s\pi N_c
\int{d^3k\thinspace d^3k^\pri\over(2\pi)^6}
{1\over\om_k\om_{k^\pri}}[2\d_{ij} - D_{ij}(\bk^\pri)]
e^{-2\om_{k^\pri}^2/\L^2}
\label{1gluon}\\
& &\quad\qquad\times\left[ {a^a_i}^\dagger(\bk) a^a_j(\bk)
+\ha\left(a^a_i(\bk) a^a_j(-\bk)
+{\rm h.c.}\right)\right]  .\nonumber
\eea
where $N_c$ is the number of colors and a quadratic
divergence has been regulated by introducing an additional
damping exponential.

The Coulomb term, Eq. (\ref{coulomb}), has a similar structure
through ${\cal O}(g^2)$.  To this order we may set
$\bdel\cdot\bD=\bdel^2$, so that
\beq
H_C = {\a_s\over 2}\int d^3x d^3y\thinspace
{\rho^a(\bx) \rho^a(\by) \over|\bx-\by|} + {\cal O}(g^3)  .
\eeq
Writing this in normal order generates two- and one-body operators.  We
shall exhibit here only those terms
involving the gluonic component of the color charge
density; the purely fermionic contributions may be found in
Ref. \cite{ss97}.

The two-body operators involving only gluons are
\bea
H_C^{(4,gg)}& & = -{\a_s\over8} f^{abc}f^{ade} \int\dkn4
\left({\om_2\om_4\over\om_1\om_3}\right)^{1/2}
\widetilde{V}(\bk_1+\bk_2)
(2\pi)^3\d^{(3)}(\sum_m\bk_m)\label{2coulomb}\\
& &\times {\bf :}
\left[ a^b_i(\bk_1) + {a^b_i}^\dagger(-\bk_1)\right]
\left[ a^c_i(\bk_2) - {a^c_i}^\dagger(-\bk_2)\right]
\left[ a^d_j(\bk_3) + {a^d_j}^\dagger(-\bk_3)\right]
\left[ a^e_j(\bk_4) - {a^e_j}^\dagger(-\bk_4)\right]
{\bf :} ,\nonumber
\eea
where
\beq
\widetilde{V}(\bk)\equiv {4\pi\over k^2} ,
\eeq
is the momentum-space Coulomb potential.  The one-body operators
obtained in normal ordering Eq. (\ref{2coulomb}) are
\bea
H_C^{(2)} = {\a_s N_c\over4} & & \int
{d^3k\thinspace d^3k^\pri\over(2\pi)^6} \widetilde{V}(\bk+\bk^\pri)
\left({\om_k^2+\om_{k^\pri}^2 \over\om_k\om_{k^\pri}}\right)
D_{ij}(\bk^\pri)
e^{-2\om_{k^\pri}^2/\L^2}
\left[ {a^a_i}^\dagger(\bk) a^a_j(\bk)\right]
\label{1coulomb}\\
+ {\a_s N_c\over8} & & \int
{d^3k\thinspace d^3k^\pri\over(2\pi)^6} \widetilde{V}(\bk+\bk^\pri)
\left({\om_{k^\pri}^2 - \om_k^2\over\om_k\om_{k^\pri}}\right)
D_{ij}(\bk^\pri) e^{-2\om_{k^\pri}^2/\L^2}
\left[a^a_i(\bk) a^a_j(\bk) + {\rm h.c.}\right]  .\nonumber
\eea
Again, quadratic divergences
have been regulated with exponential
cut-offs.

There are also mixed fermion-gluon two-body operators, given by
\bea
H_C^{(4,qg)} &=& -{i\a_s \over2}\sum_{s,s^\pri} \int
{d^3k\thinspace d^3k^\pri\over(2\pi)^6}
{d^3q\thinspace d^3q^\pri\over(2\pi)^6}
\thinspace f^{abc} T^a_{\a\b}
\left({\om_{\bq^\pri}\over\om_{\bq}}\right)^{1/2}
\widetilde{V}(\bq+\bq^\pri)\nonumber\\
&\times & (2\pi)^3\d^{(3)}(-\bk+\bk^\pri+\bq+\bq^\pri)
\; {\bf :}\left[ a_i^b(\bq)+{a_i^b}^\dagger(-\bq) \right]
\left[ a_i^c(\bq^\pri)-{a_i^c}^\dagger(-\bq^\pri) \right]\\
&\times &
\Bigl(\left[u^\dagger(\bk,s) u(\bk^\pri,s^\pri)\right]
b^\dagger_\a (\bk,s) b_\b(\bk^\pri,s^\pri)
+ \left[v^\dagger(-\bk,s) v(-\bk^\pri,s^\pri)\right]
d_\a(-\bk,s) d^\dagger_\b(-\bk^\pri,s^\pri) \nonumber\\
& & + \left[u^\dagger(\bk,s) v(-\bk^\pri,s^\pri)\right]
b^\dagger_\a(\bk,s) d^\dagger_\b(-\bk^\pri,s^\pri)
+ \left[v^\dagger(-\bk,s) u(\bk^\pri,s^\pri)\right]
d_\a(-\bk,s) b_\b(\bk^\pri,s^\pri)\Bigr){\bf :}. \nonumber
\eea
No tadpoles arise from normal ordering this operator since
$T^a$ has zero trace  and the
$f^{abc}$ are antisymmetric.

Finally there is the quark-gluon coupling term $H_{qg}$
[Eq. (\ref{hqg})]
\bea
H_{qg} =& & -g\sum_{s,s^\pri}\int {d^3k\thinspace
d^3k^\pri\thinspace d^3q\over(2\pi)^9}
\thinspace{T^a_{\a\b}\over \sqrt{2\om_q}}\thinspace
(2\pi)^3\d^{(3)}(-\bk+\bk^\pri+\bq)
\thinspace{\bf :}\left[a_i^a(\bq)+{a_i^a}^\dagger(-\bq)\right] \\
\times\;
\Bigl(& &\left[u^\dagger(\bk,s)\a_i u(\bk^\pri,s^\pri)\right]
b^\dagger_\a (\bk,s) b_\b(\bk^\pri,s^\pri)
+ \left[v^\dagger(-\bk,s)\a_i v(-\bk^\pri,s^\pri)\right]
d_\a(-\bk,s) d^\dagger_\b(-\bk^\pri,s^\pri) \nonumber\\
+& &\; \left[u^\dagger(\bk,s)\a_i v(-\bk^\pri,s^\pri)\right]
b^\dagger_\a(\bk,s) d^\dagger_\b(-\bk^\pri,s^\pri)
+ \left[v^\dagger(-\bk,s)\a_i u(\bk^\pri,s^\pri)\right]
d_\a(-\bk,s) b_\b(\bk^\pri,s^\pri)\Bigr){\bf :}	.\nonumber
\eea

%

\end{document}